\documentclass[aps,11pt,nofootinbib,groupedaddress,preprintnumbers,superscriptaddress]{revtex4-2}

\usepackage{amssymb, amsfonts, amsmath, bm}
\usepackage[
]{graphicx}
\usepackage{color}

\usepackage[
]{hyperref}
\hypersetup{%
 setpagesize=false,
 bookmarksnumbered=true,%
 bookmarksopen=true,%
 colorlinks=true,%
 linkcolor=blue,%
 citecolor=red}

\begin{document}
\title{Stable circular orbits in caged black hole spacetimes}
\author{Takahisa Igata}
\email{igata@post.kek.jp}
\affiliation{KEK Theory Center, 
Institute of Particle and Nuclear Studies, 
High Energy Accelerator Research Organization, Tsukuba 305-0801, Japan}
\author{Shinya Tomizawa}
\email{tomizawa@toyota-ti.ac.jp}
\affiliation{
Mathematical Physics Laboratory, 
Toyota Technological Institute, Nagoya 468-8511, Japan}
\date{\today}
\preprint{KEK-TH-2293}
\preprint{KEK-Cosmo-0271}
\preprint{TTI-MATHPHYS-3}
 
\begin{abstract}
We consider the motion of massive and massless particles in a five-dimensional spacetime 
with a compactified extra-dimensional space where a black hole is localized, 
i.e., a caged black hole spacetime. 
We show the existence of circular orbits and reveal their sequences and stability. 
In the asymptotic region, stable circular orbits always exist, which implies that four-dimensional gravity is more dominant because of the small extra-dimensional space. 
In the vicinity of a black hole, they do not exist because the effect of compactification is no longer effective.
We also clarify the dependence of the sequences of circular orbits on the size of the extra-dimensional space by determining the appearance of the 
innermost stable circular orbit and the last circular orbit (i.e., the unstable photon circular orbit). 
\end{abstract}
\maketitle

\section{Introduction}
\label{sec:1}
We naively perceive our world as a $(3+1)$-dimensional spacetime.
However, in the context of unified theories, 
a higher-dimensional model of the Universe that adds an extra-dimensional space 
to the four-dimensional (4D) spacetime has been studied 
for a long time~\cite{Kaluza:1921tu, Klein:1926tv}.
In this research background, higher-dimensional black holes 
have been actively studied as a field 
to find the various properties of higher-dimensional spacetime and 
gravity~\cite{Emparan:2008eg}.
In understanding the nature of higher-dimensional black hole spacetimes, 
it is essential to consider test particle dynamics and compare it to that in 4D.
As a first step, many studies were carried out on the motion of particles 
in an asymptotically flat higher-dimensional black hole spacetime 
with a single spherical horizon~\cite{Tangherlini:1963bw,Myers:1986un}. 
They revealed one of the most distinctive differences 
from 4D due to the dimensional dependence of gravity, 
the absence of the stable circular 
orbit~\cite{Hackmann:2008tu,Frolov:2003en,Diemer:2014lba,Cardoso:2008bp}.%
\footnote{Note that stable stationary/bound orbits can exist 
in the ultraspinning regime of the Myers-Perry black holes 
in more than six dimensions~\cite{Igata:2014xca}.}
As a result, the features of 4D gravity are gradually highlighted. 
Furthermore, since the uniqueness theorem does not hold 
for higher-dimensional black holes as in 4D~\cite{Hollands:2007aj, Hollands:2012xy}, 
and they can have a nonspherical 
horizon (e.g., ring and lens spaces~\cite{Emparan:2001wn,Kunduri:2014kja,Tomizawa:2016kjh}), 
various particle dynamics depending on the horizon topology 
can also occur in higher dimensions. 
Indeed, stable circular/bound orbits appear 
in the five-dimensional (5D) black ring 
spacetime~\cite{Hoskisson:2007zk,Igata:2010ye,Grunau:2012ai,Igata:2013be,Igata:2020vdb,Igata:2020dow}. 
It was recently shown that stable circular/bound orbits also exist 
in the 5D supersymmetric 
black lens spacetimes~\cite{Tomizawa:2019egx,Tomizawa:2020mvw}.

The next step is to consider black hole spacetimes 
that model how we cannot observe an extra-dimensional space.
One of the possible mechanisms to explain such inability is 
the compactification of the extra-dimensional space.
Black hole spacetimes that incorporate this mechanism are 
called Kaluza-Klein black holes, 
and many solutions of this class have been found in the higher-dimensional 
Einstein gravity so far (see, e.g., Ref.~\cite{Tomizawa:2011mc} and references therein).
Focusing on 5D Kaluza-Klein black holes, we can classify them into two major classes.
One is the class in which the horizon is spread out over the whole extra-dimensional space.
The other is the class in which the horizon is localized 
in a certain portion of the extra-dimensional space, 
the so-called caged Kaluza-Klein 
black holes~\cite{Myers:1986rx,Kol:2003if,Harmark:2003yz,Suzuki:2012av}.
How the existence of a compact extra dimension has nontrivial effects on particle dynamics 
is an important and nontrivial question.
Particle dynamics in the former class has been well 
studied~\cite{Lim:1992,Kalligas:1994vf,Liu:2000zq,Matsuno:2009nz,Long:2019nox} 
because of its relatively higher symmetry. 
On the other hand, particle dynamics in the latter class has not been well 
investigated because of its relatively lower symmetry.

However, it was recently shown that stable circular orbits exist 
by the many-body effect of black holes if the separation 
between the horizons is large enough in a 5D multi--black hole spacetime~\cite{Igata:2020vlx}.
Since a caged black hole can be identified with an infinite number of black holes 
localized in a one-dimensional direction, 
such many-body effects can be expected to be inherited to particle dynamics 
in the caged black hole spacetime.
The purpose of this paper is to reveal the effects of an extra dimension 
through the dynamics of particles moving in 
the caged black hole backgrounds~\cite{Myers:1986rx}. 
In the region sufficiently far from the black hole, the particle dynamics is like 4D, 
while in the near horizon, the effect that a black hole is localized 
in a compactified dimension appears more effectively.

This paper is organized as follows. In Sec.~\ref{sec:2}, we introduce 
a 5D caged black hole spacetime and 
formulate conditions for stable/unstable circular orbits in the spacetime. 
In Sec.~\ref{sec:3}, we clarify the dependence of sequences of 
circular orbits on the size of extra-dimensional space. 
Section~\ref{sec:4} is devoted to a summary and discussions. 
Throughout this paper, we use units in which $G=1$ and $c=1$, 
where $G$ is the 5D Newton constant and $c$ is the speed of light.

\section{Formulation}
\label{sec:2}
We shortly review the caged black hole spacetime given in Ref.~\cite{Myers:1986rx}. 
We begin by considering the metric and gauge field 
in the 5D Majumdar-Papapetrou geometry,
\begin{align}
&g_{\mu\nu}\:\!\mathrm{d}x^\mu\:\!\mathrm{d}x^\nu
=-U^{-2}(\bm{x})\:\!\mathrm{d}t^2
+U(\bm{x}) \:\! \mathrm{d}\bm{x}\cdot \mathrm{d}\bm{x},
\\
&A_\mu\:\!\mathrm{d}x^\mu=-\frac{\sqrt{3}}{2} U^{-1}(\bm{x}) \:\!\mathrm{d}t,
\end{align}
where $t$ is the global Killing time, and $\bm{x}$ denotes spatial coordinates, 
and $\mathrm{d}\bm{x}\cdot \mathrm{d}\bm{x}$ is the metric in 
the 4D Euclidean space $\mathbb{E}^4$.
For these ansatz, 
the only nontrivial components in the field equations are 
the $(t, t)$ component of the Einstein equation and 
the $t$ component of the Maxwell equation,%
\footnote{The field equations are derived from the 5D Einstein-Maxwell theory, 
\begin{align}
S=\int \mathrm{d}^5 x \:\!\sqrt{-g}\:\!(R-F_{\mu\nu} F^{\mu\nu}),
\end{align}
where $R$ is the Ricci tensor and 
$F_{\mu\nu}$ is the field strength of the gauge field.} 
both of which are equivalent to the Laplace equation in $\mathbb{E}^4$, 
\begin{align}
\label{eq:Laplace}
\Delta_{\mathbb{E}^4} U=0.
\end{align}
Let us introduce the coordinates $\bm{x}=(\rho, \theta, \phi, w)$ 
in which the Euclidean metric takes the form
\begin{align}
\mathrm{d}\bm{x}\cdot \mathrm{d}\bm{x}
=\mathrm{d}\rho^2+\rho^2 (\mathrm{d}\theta^2+\sin^2\theta\:\!
\mathrm{d}\phi^2)+\mathrm{d}w^2.
\end{align}
Consider a solution $U$ of Eq.~\eqref{eq:Laplace} 
for an infinite number of point sources of mass scale $\mu$ on the $w$-axis 
with equal spacing $a=2\pi\ell$, 
\begin{align}
U&=1+\sum_{n=-\infty}^\infty \frac{\mu}{\rho^2+(w+n a)^2}
\\
&=1+\frac{\pi \mu}{a \rho}\frac{\sinh (\pi \rho/a) \cosh (\pi \rho/a)}{\sin^2 (\pi w/a) +\sinh^2 ( \pi \rho/a)}
\\
&=1+\frac{\mu}{2\ell \rho} \frac{\sin h (\rho/\ell)}{\cosh(\rho/\ell)-\cos(w/\ell)},
\end{align}
where the dimension of $\mu$ is length squared even in ordinary units. 
This function has reflection symmetry under $w \to -w$. 
Furthermore, $U$ is periodic in $w$ with period $a$, 
and therefore, we may periodically identify the spacetime in the $w$ direction. 
As a result, we have a spacetime where a single black hole 
with $S^3$ horizon topology is localized in a compactified extra dimension, 
which is referred to as the caged black hole spacetime. 
Thus, the parameter $\ell$ corresponds to the 
radius of the $S^1$-compactified extra-dimensional space. 
We only focus on the range $-\pi \ell <w\leq \pi \ell$ in what follows.

We check the structure of the gravitational field of the caged black hole spacetime 
at several scales through the asymptotic shape of $U$.
It is useful to gain an intuition for the dynamics of particles.
In the region where $\rho, w \ll a$,
the function $U$ is expanded as 
\begin{align}
\label{eq:Ushort}
U=1+\frac{\mu}{r^2}+ \frac{\pi^2}{3}\frac{\mu}{a^2} + O(\rho^2/a^2, w^2/a^2),
\end{align}
where $r^2=\rho^2+w^2$.
The second term corresponds to the monopole term appearing 
in the case of 5D asymptotically flat black holes. 
The third term 
is contributions to the potential in the short-range from all the other image sources.%
\footnote{
\begin{align}
\sum_{n=1}^{\infty}\frac{1}{n^2}=\frac{\pi^2}{6}.
\end{align}}
Therefore, we can expect that 
the particle dynamics in this region is the same as that 
in a 5D asymptotically flat black hole spacetime.

In the region where $\rho \gg a$, the function $U$ is expanded as
\begin{align}
\label{eq:asymU}
U
=1+\frac{\mu}{2\:\!\ell \rho}+\frac{\mu}{\ell \rho}e^{-\rho/\ell}\cos (w/\ell) +\cdots.
\end{align}
Note that the third and subsequent terms are exponentially suppressed, 
and thus, the metric reduces to a black string (ring). 
The power of $\rho$ in the second term implies that 
test particles in the asymptotic region feel gravitational force 
as in 4D asymptotically flat black hole spacetimes.

We consider the dynamics of a freely falling particle 
with unit/zero mass in the caged black hole spacetime. 
Let $p_\mu$ be the canonical momenta conjugate with coordinate variables of a particle. 
The Hamiltonian of affinely parametrized geodesics is given by
\begin{align}
H=\frac{1}{2} g^{\mu\nu}p_\mu p_\nu
=
-\frac{U^2}{2} E^2 +\frac{1}{2\:\!U} \left(
p_w^2+p_\rho^2+\frac{L^2}{\rho^2} 
\right),
\end{align}
where $E=-p_t$ is constant particle energy, and 
\begin{align}
L^2=p_\theta^2+\frac{p_\phi^2}{\sin^2\theta}
\end{align}
is a constant associated with the $S^2$ rotational symmetry. 
From the on-shell condition, $g^{\mu\nu}p_\mu p_\nu=-\kappa$, 
where $\kappa$ is particle mass squared, 
we obtain the constraint equation 
\begin{align}
\label{eq:eneq}
&U^{-1} (\dot{\rho}^2+\dot{w}^2)+V=E^2,
\\
\label{eq:Vdef}
&V(\rho, w; L^2)=\frac{L^2}{\rho^2 U^3}+\frac{\kappa}{U^2},
\end{align}
where the dots denote the derivatives with respect to an affine parameter. 
We call $V$ the effective potential of the two-dimensional (2D) dynamics 
in the $(\rho, w)$ plane.

We focus on stationary orbits of particles with $\kappa=1$, 
in which $\rho$ and $w$ remain constant. 
Note that all of the stationary orbits are circular 
because of the $S^2$ rotational symmetry. 
The conditions of the stationary orbits for $V$ and
$V_i=\partial_i V$ ($i=w, \rho$) are written as
\begin{align}
\label{eq:Vw}
V_w&=-\frac{2\:\!U_w}{U^3}\left(
1+\frac{3}{2}\frac{L^2}{\rho^2 U}
\right)=0,
\\
\label{eq:Vrho}
V_\rho&=-\frac{2 L^2}{\rho^3 U^3}
-\frac{2\:\!U_\rho}{U^3}\left(
1+\frac{3}{2}\frac{L^2}{\rho^2 U}
\right)=0,
\\
\label{eq:V=E2}
V&=E^2,
\end{align}
where the explicit forms of $U_i=\partial_i U$ ($i=w, \rho$) are given by
\begin{align}
U_w&=-\frac{\mu}{2\:\!\ell^2 \rho}\frac{\sin(w/\ell) \sinh (\rho/\ell)}{ \left[\:\!
\cos (w/\ell)-\cosh (\rho/\ell)
\:\!\right]^2}, 
\\
U_\rho&=\frac{\mu}{2\ell^2\rho^2}\frac{\rho \left[\:\!
1-\cos(w/\ell) \cosh(\rho/\ell)
\:\!\right]+\ell \sinh(\rho/\ell) \left[\:\!
\cos (w/\ell)-\cosh(\rho/\ell)
\:\!\right]}{\left[\:\!\cos(w/\ell)-\cosh(\rho/\ell)\:\!\right]^2}.
\end{align}
The condition~\eqref{eq:Vw} leads to $U_w=0$, i.e., 
\begin{align}
\label{eq:Dzeta0}
w=0, \ \pi \ell.
\end{align}
These correspond to the fixed points of the reflection symmetry of $U$. 
Furthermore, solving the conditions~\eqref{eq:Vrho} and \eqref{eq:V=E2} 
for $L^2$ and $E^2$, we obtain
\begin{align}
L^2&=L_0^2(\rho, w)
:=-\frac{2\rho^3 U U_\rho}{f},
\\
E^2&=E_0^2(\rho, w):=V(\rho, w; L_0^2)=\frac{2 U+\rho \:\!U_\rho}{fU^2},
\end{align}
where 
\begin{align}
f(\rho, w)=2\:\!U+3\rho\:\! U_\rho.
\end{align}
These must be non-negative to find circular orbits on $w=0$ or $\pi \ell$. 
Therefore, we can represent the sequence of circular orbits on the $(\rho, w)$ plane as 
\begin{align}
\gamma_0=\left\{
(\rho, w) \:\!|\:\! (w=0\ \mathrm{or} \ w=\pi \ell), L_0^2 \geq 0
\right\},
\end{align}
where we have used the fact that $L^2\geq0$ always means $E^2>0$ 
because of Eqs.~\eqref{eq:eneq} and \eqref{eq:Vdef}. 
The explicit forms of $L_0^2$ and $E_0^2$ on $\gamma_0$ are given by
\begin{align}
L_0^2(\rho, \Theta(\sigma)\:\! \pi \ell)
&=\frac{\mu \rho}{\ell}
\frac{\left[\:\!
-\sigma \rho+\ell \sinh (\rho/\ell)
\:\!\right]\left[\:\!
2\ell \rho +\mu \left(\tanh\left[\:\!
\rho/(2\ell)
\:\!\right]\right)^{\sigma}
\:\!\right]}{4\ell^2\rho \cosh(\rho/\ell)-\mu \ell \sinh (\rho/\ell)+\sigma (3\mu+4\ell^2)\rho},
\\
E_0^2(\rho, \Theta(\sigma)\:\!\pi \ell)
&= \frac{4\ell^2 \rho^2 \left(
\mu \left[\:\!
\sigma \rho+\ell \sinh (\rho/\ell)
\:\!\right]+4\ell^2\rho \left[\:\!
\sigma+\cosh (\rho/\ell)
\:\!\right]\right)}{\left[\:\!
2\ell \rho+\mu \left(
\tanh \left[\:\!
\rho/(2\ell)
\:\!\right]
\right)^{\sigma}
\:\!\right]^2 \left[\:\!
4\ell^2\rho \cosh (\rho/\ell)
-\mu \ell \sinh (\rho/\ell)
+\sigma (3\mu+4\ell^2) \rho
\:\!\right]},
\end{align}
respectively, where $\sigma=\pm1$, and $\Theta(\sigma)$ denotes the Heaviside step function, and we have used 
\begin{align}
f(\rho, \Theta(\sigma) \:\!\pi \ell)
=2-\frac{\mu}{ 4\ell^2\rho} \frac{- 3\sigma\rho+\ell \sinh (\rho/\ell)}{\Theta(\sigma)\cosh^2 \left[\:\!
\rho/(2\ell)
\:\!\right]+\Theta(-\sigma) \sinh^2\left[\:\!\rho/(2\ell)\:\!\right]}.
\end{align}
The sign of $f(\rho, \Theta(\sigma) \:\!\pi \ell)$ determines the signs of 
$L_0^2(\rho, \Theta(\sigma)\:\! \pi \ell)
$ and $E_0^2(\rho, \Theta(\sigma)\:\!\pi \ell)$. 
They diverge at $f(\rho, \Theta(\sigma) \:\!\pi \ell)=0$.

We further classify $\gamma_0$ by imposing stability conditions for circular orbits. 
Let $(V_{ij})$ be the Hessian matrix of $V$ on the 2D flat space with $\delta_{ij}\:\!\mathrm{d}x^i\:\!\mathrm{d}x^j=\mathrm{d}\rho^2+\mathrm{d}w^2$, where $V_{ij}=\partial_j \partial_i V$ ($i, j=\rho, w$). 
Let $h$ and $k$ be the determinant and the trace of $(V_{ij})$ on $\mathbb{E}^2$, i.e., 
$h(\rho, w; L^2)=\mathrm{det}(V_{ij})$ and $k(\rho, w; L^2)=\mathrm{tr}(V_{ij})$, respectively.
Since we analyze the particle dynamics on a 2D reduced space, of which metric is $\tilde{\gamma}_{ij}=\Omega^2 \delta_{ij}=U^{-1} \delta_{ij}$ [see Eq.~\eqref{eq:eneq}],
then the stability of circular orbits should be determined on the basis of the Hessian matrix $(\tilde{V}_{ij})=(\tilde{\nabla}_j \tilde{\nabla}_i V)$ in the 2D conformally flat space, where $\tilde{\nabla}_i$ is the covariant derivative associated with $\tilde{\gamma}_{ij}$. 
Focus on the relation between $\tilde{V}_{ij}$ and $V_{ij}$,
\begin{align}
\tilde{V}_{ij}
=V_{ij}-\Omega^{-1}
(2 V_{(i} \Omega_{j)}-\delta_{ij}\delta^{kl}V_k \Omega_l),
\end{align}
where $\Omega_i=\partial_i \Omega$. 
Note that $\tilde{V}_{ij}
=V_{ij}$ on $\gamma_0$ because 
$V_i=0$ there.
Furthermore, on $\gamma_0$, the trace and determinant of $(\tilde{V}_{ij})$ 
coincide with $U k $ and 
$U^2 h$, respectively. Therefore, we can use $k$ and $h$ 
to determine the signs of the trace and determinant 
of $(\tilde{V}_{ij})$, respectively. 
In terms of them, we define the region $D$ such that 
\begin{align}
D=\left\{\:\!(\rho, w)\:\!|\:\!
h_0>0, k_0>0, L_0^2>0\:\!\right\},
\end{align}
where $h_0$ and $k_0$ are defined as 
\begin{align}
h_0(\rho, w)&:=h(\rho, w; L_0^2)\big|_{U_w=0}=\frac{-16\rho\:\! U^2 U_{w\rho}^2+8 \:\!U_{ww}\left[\:\!
6\rho\:\! U U_\rho^2+3\:\!\rho^2 U_\rho^3+2\:\!U^2(3\:\!U_\rho+\rho\:\! U_{\rho\rho})
\:\!\right]}{\rho \:\!U^6 f^2},
\\
k_0(\rho, w)&:=k(\rho, w; L_0^2)\big|_{U_w=0}=-\frac{2}{\rho \:\!U^4} \frac{6\rho \:\!U U^2_{\rho} +3\rho^2 U_\rho^3+2\:\!U^2(3\:\!U_\rho+\rho\:\! U_{\rho\rho}+\rho \:\!U_{ww})}{f}.
\end{align}
The restriction that $U_w=0$ means that 
the terms proportional to $U_w$ have been removed. 
As a result, the part of $\gamma_0$ overlapped by $D$ is the sequence of stable circular orbits, and its boundaries correspond to the marginally stable circular orbits. On the other hand, the part of $\gamma_0$ without overlap with $D$ is the sequence of unstable circular orbits.

\section{Circular orbits in the caged black hole spacetimes}
\label{sec:3}
\begin{figure}[t]
\centering
\includegraphics[width=16cm,clip]{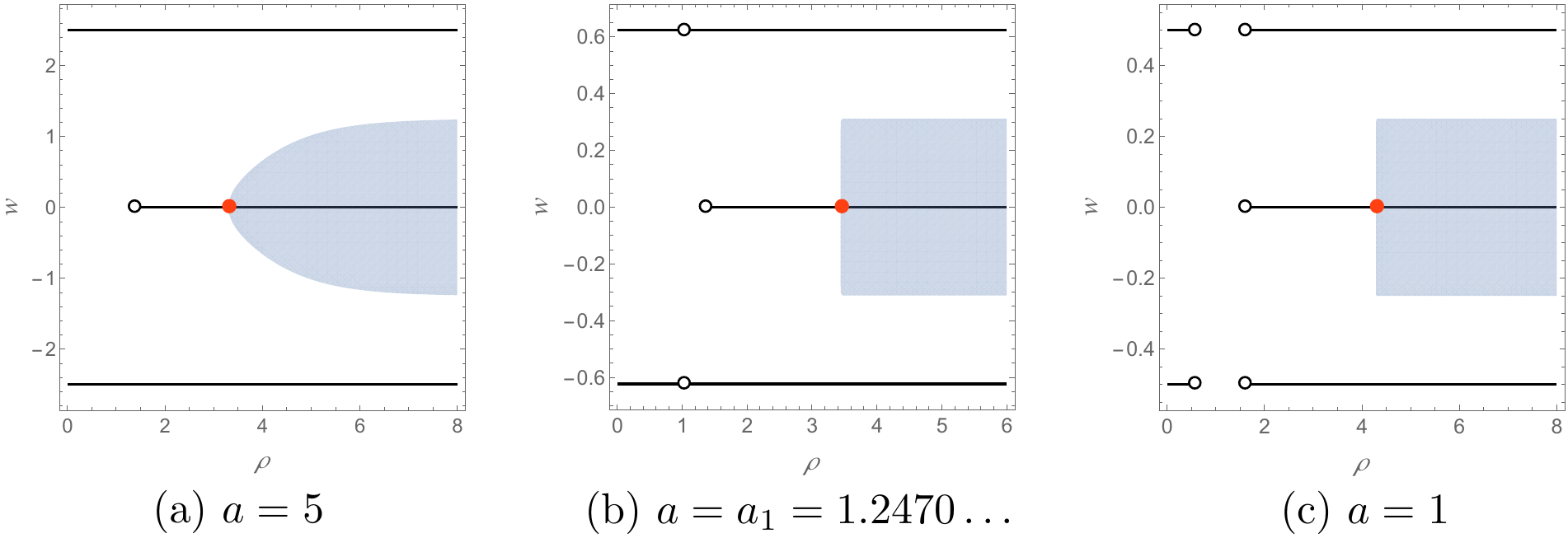}
 \caption{
Sequences of stable/unstable circular orbits for several sizes of the extra dimension. 
We use units in which $\mu=1$. 
Black solid lines show $\gamma_0$, sequences of circular orbits, 
and blue shaded regions show $D$, 
inside which circular orbits are stable. 
Red dots denote the ISCOs, and white circles denote UPCOs. 
 }
 \label{fig:CO}
\end{figure}

We consider circular orbits in the 5D caged black hole spacetimes 
by using the quantities introduced in the previous section. 
First, we illustrate typical sequences of circular orbits by comparing the size of the extra dimension $a$ and the mass parameter $\mu$.
We use units in which $\mu=1$ in what follows. 
Figure~\ref{fig:CO}(a) shows the case $a=5$, 
typical sequences of circular orbits for $a\gg 1$. 
The black solid lines are $\gamma_0$, and the blue shaded region is $D$. 
The part of $\gamma_0$ overlapped by $D$ appears on $w=0$,
a sequence of stable circular orbits, 
which extends from the innermost stable circular orbit (ISCO) $\rho=\rho_{\mathrm{I}}$~(indicated by a red dot) to infinity. 
The energy and squared angular momentum, $E_0$ and $L_0^2$, 
decrease monotonically with $\rho$ 
(i.e., $\mathrm{d}E_0(\rho, 0)/\mathrm{d}\rho\geq 0$ 
and $\mathrm{d}L_0^2(\rho, 0)/\mathrm{d}\rho\geq 0$) in the range $\rho_{\mathrm{I}}\leq \rho<\infty$. 
Each of them takes a local minimum value at the ISCO, where $h_0(\rho_{\mathrm{I}}, 0)=0$ also holds. 
On the other hand, a sequence of unstable circular orbits appears 
on the segment of $\gamma_0$ between the ISCO and 
the last circular orbit $\rho=\rho_{\mathrm{p}}$ (denoted by a white circle). 
In this range, the energy and squared angular momentum satisfy $\mathrm{d}E_0(\rho, 0)/\mathrm{d}\rho<0$ 
and $\mathrm{d}L_0^2(\rho, 0)/\mathrm{d}\rho<0$, respectively, 
and diverge in the limit to the white circle.
The last circular orbit on $w=0$ is justified as an unstable photon circular orbit~(UPCO)%
\footnote{The conventional term ``photon" is used to describe the unstable circular orbit of a massless particle.}
because the ratio $L_0/E_0$ is still finite even in the limit. 
We also find a sequence of unstable circular orbits on $w=\pi \ell$. 
Next, let us see the case where $a$ takes a smaller value. 
Figure~\ref{fig:CO}(b) shows the case $a=a_1$, where 
$E_0$ and $L_0^2$ on $w=\pi \ell$ diverge 
at a radius $\rho=\rho_1$ (white circle), where 
\begin{align}
a_1&=1.2470\ldots,
\\
\rho_1&=1.0129\ldots.
\end{align}
It corresponds to a UPCO on $w=\pi \ell$. 
Figure~\ref{fig:CO}(c) shows the case $a=1$, 
typical sequences of circular orbits for $a\lesssim 1$. 
Even in the range, 
we can see a sequence of stable circular orbits 
between infinity and the ISCO on $w=0$ 
and can also see a sequence of unstable circular orbits 
between the ISCO and the last circular orbit (i.e., the UPCO). 
The difference appears in sequences on $w=\pi \ell$, which separate into two 
pieces. Each end point of the sequences corresponds to a UPCO. 

\begin{figure}[t]
\centering
\includegraphics[width=15cm,clip]{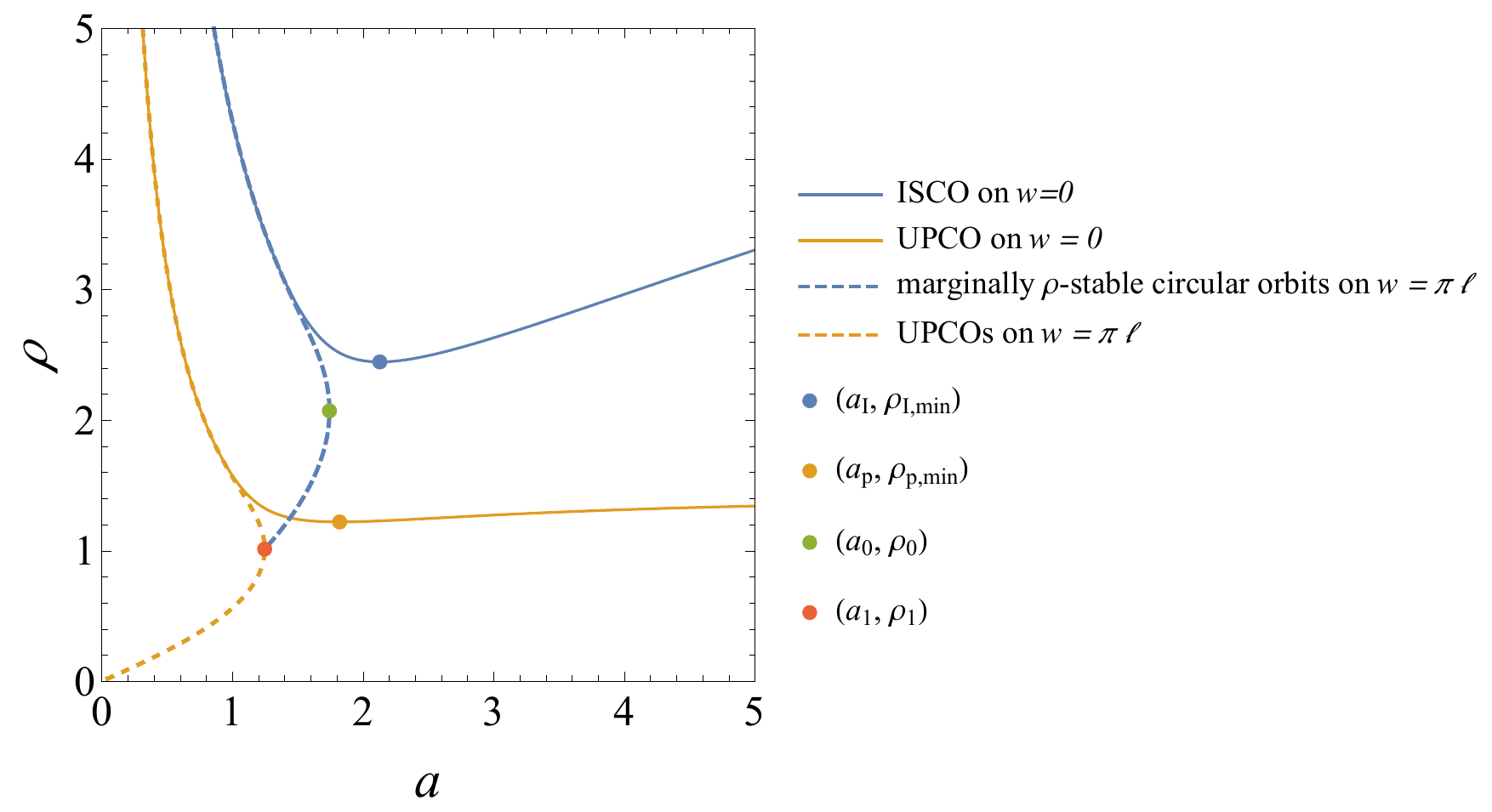}
 \caption{Dependence of the radii of 
 the ISCO, UPCOs, and marginally $\rho$-stable circular orbits 
 on the size of the extra dimension, $a$. 
 We use units in which $\mu=1$. 
 Blue solid curve denotes the radius of the ISCO on $w=0$. 
 Blue dashed curve denotes a pair of radii of 
 marginally $\rho$-stable circular orbits 
 on $w=\pi \ell$. 
 Orange solid and dashed curves show 
 the radii of UPCOs on $w=0$ and $w=\pi \ell$, respectively. 
 }
 \label{fig:ISCO}
\end{figure}

Figure~\ref{fig:ISCO} shows the dependence of some characteristic orbital radii on $a$. 
The blue solid curve shows the ISCO radius $\rho=\rho_{\mathrm{I}}$ 
as a function of $a$, 
which is determined by $h_0(\rho_{\mathrm{I}}, 0)=0$.
For $a>a_{\mathrm{I}}$, the radius $\rho_{\mathrm{I}}$ monotonically decreases
as $a$ decreases, whereas for $a<a_{\mathrm{I}}$, 
it monotonically increases as $a$ decreases, where 
\begin{align}
a_{\mathrm{I}}=2.1286\ldots.
\end{align}
Hence, at $a=a_{\mathrm{I}}$, the ISCO radius takes the minimum value (see the blue dot)
\begin{align}
\rho_{\mathrm{I}, \mathrm{min}}=2.4465\ldots.
\end{align}
The orange solid curve shows the last circular orbit radius $\rho=\rho_{\mathrm{p}}$ (or equivalently, the UPCO radius) as a function of $a$, which is determined by $f(\rho_{\mathrm{p}}, 0)=0$. For $a>a_{\mathrm{p}}$, 
the radius $\rho_{\mathrm{p}}$ monotonically decreases as $a$ decreases, whereas for $a<a_{\mathrm{p}}$, 
it monotonically increases as $a$ decreases, where 
\begin{align}
a_{\mathrm{p}}=1.8206\ldots.
\end{align}
At $a=a_{\mathrm{p}}$, the radius of the UPCO on $w=0$ takes the minimum value (see the orange dot)
\begin{align}
\rho_{\mathrm{p}, \mathrm{min}}=1.2210\ldots.
\end{align}

The blue dashed curve shows a pair of circular orbit radii on $w=\pi \ell$ 
that are marginally stable against small perturbations only in the $\rho$ direction, 
which are determined by $V_{\rho\rho}(\rho, \pi \ell; L_0^2(\rho, \pi \ell))=0$. 
We call them marginally $\rho$-stable circular orbits. 
The outer radius $\rho\geq \rho_0$
appears only in the range $0<a\leq a_0$, where 
\begin{align}
a_0&=1.7430\ldots, 
\\
\rho_0&=2.0717\ldots, 
\end{align}
and increases as $a$ decreases. 
The inner radius $\rho\leq \rho_0$ appears only in the range 
$a_1<a<a_0$ and decreases with $a$ and disappears at $a=a_1$. 
The orange dashed curve shows a pair of the radii of UPCOs on $w=\pi \ell$. 
The inner radius decreases with $a$ and finally goes to zero in the limit $a\to 0$. 
The outer radius increases as $a$ decreases. 
There are no circular orbits between these radii. 
In the enclosed region by the blue and orange dashed curves, 
the circular orbits that are unstable in all directions appear on $w=\pi \ell$, 
and the energy and squared angular momentum satisfy 
$\mathrm{d}E_0(\rho, \pi \ell)/\mathrm{d}\rho<0$ and 
$\mathrm{d}L_0^2(\rho, \pi \ell)/\mathrm{d}\rho<0$, respectively. 
In the region to the right of all the dashed curves, 
$\rho$-stable circular orbits appear on $w=\pi \ell$, and 
the energy and squared angular momentum satisfy 
$\mathrm{d}E_0(\rho, \pi \ell)/\mathrm{d}\rho>0$ and 
$\mathrm{d}L_0^2(\rho, \pi \ell)/\mathrm{d}\rho>0$, respectively.

Consider the qualitative behaviors of particle dynamics in the asymptotic analysis of $V$.
We restore $\mu$ in the following discussions. 
We can see that $V$ in the asymptotic region $\rho\gg a$ behaves like the 
effective potential of a 4D asymptotically flat black hole spacetime, 
as is expected from Eq.~\eqref{eq:asymU}, as 
\begin{align}
\label{eq:Vasym}
V=1-\frac{\mu}{\ell \rho}+\frac{L^2}{\rho^2}
-\frac{3\mu L^2}{2\ell \rho^3}+O(\ell e^{-\rho/\ell}/\rho).
\end{align}
The second term implies that the gravitational mass of the black hole, 
as perceived by the particle, 
is proportional to 
$M_{\mathrm{grav}}=\mu c^2/(2\:\!\ell G_4)$, 
where we have restored the speed of light $c$ and the 4D Newton constant $G_4=G/a$. 
Hence, the mass increases as $\ell$ decreases. 
Evaluating the ISCO radius up to this order, 
we find $\rho_{\mathrm{I}}=(9/2)(\mu/\ell)=(9/2)r_{\mathrm{g}}$,
where $r_{\mathrm{g}}=2G_4 M_{\mathrm{grav}}/c^2$ is the Schwarzschild radius. 
Furthermore, as can be seen from the fact that 
the leading terms in Eq.~\eqref{eq:Vasym} are independent of $w$, 
gravitational force in the $\rho$ direction is dominant in the asymptotic region. 
As a result, the ISCO and the marginally $\rho$-stable circular orbit radii there 
increase as $a$ decreases, and they must have the same value regardless of $w$, i.e., 
in this region, the solid and dashed blue curves in Fig.~\ref{fig:ISCO} coincide with each other.
The same behavior can be seen for UPCOs, i.e., 
the solid and dashed orange curves coincide with each other in this region.

We find from Eq.~\eqref{eq:Ushort} that $V$ in the range $\rho, w \ll a$
behaves like the effective potential of a 5D asymptotically flat black hole spacetime as 
\begin{align}
\label{eq:Vnear}
V=1-\frac{2\pi^2}{3}\frac{\mu}{a^2}-\frac{2\mu}{\rho^2+w^2}
+\left(1-\frac{\pi^2 \mu}{a^2}\right)\frac{L^2}{\rho^2}
-\frac{3\mu L^2}{\rho^2 (\rho^2+w^2)}+O(\rho^2/a^2, w^2/a^2).
\end{align}
The third term corresponds to a 5D gravitational potential. 
In particular, on $w=0$, the potential $V$ of Eq.~\eqref{eq:Vnear} reduces to 
\begin{align}
V(\rho, 0)=1-\frac{2\pi^2}{3}\frac{\mu}{a^2}
+\frac{(1-\pi^2\mu/a^2) L^2-2\mu}{\rho^2}
-\frac{3\mu L^2}{\rho^4}+O(\rho^2/a^2). 
\end{align}
Thus, we find that there are no circular orbits in this range 
because the third and fourth terms cannot make a potential well.

\section{Summary and discussions}
\label{sec:4}

We have considered sequences of circular orbits for massive and massless particles 
in the 5D caged black hole spacetime, 
in which a black hole is localized in the extra-dimensional space. 
We have given a systematic way to find stationary orbits (i.e., circular orbits) 
and a prescription to determine whether they are stable or unstable.
Using these, we have identified a typical sequence of circular orbits 
for each size of extra-dimensional space and have specified the part 
where it shows stable behavior.

We have found that stable circular orbits exist in the asymptotic region regardless of 
the scales of the extra dimension and the black hole mass. 
It implies that the localization effect of the black hole 
in the extra-dimensional space does not appear in the region far from the black hole. 
The existence of stable circular orbits in such an asymptotic region is analogous to 
the case of a 4D asymptotically flat black hole spacetime, 
rather than a 5D asymptotically flat black hole spacetime with a spherical horizon. 
In other words, we can interpret the effect of the compactification of 
the extra-dimensional space in the asymptotic region as reproducing effective 4D gravity.
As mentioned in the Introduction, we can also interpret this phenomenon 
as a consequence of the many-body effect 
due to the infinite images of a black hole~\cite{Igata:2020vlx}.
On the other hand, 
in the region closer to the black hole than the size of the extra dimension, 
stable circular orbits do not appear 
because 5D gravity of the asymptotically flat black hole spacetime dominates 
due to the suppression of the compactification effect.
In the intermediate region between these two, 
the sequence of stable circular orbits reaches the ISCO 
and switches to the unstable circular orbits, 
and finally, it terminates in the last circular orbit (i.e., the UPCO).
This behavior does not qualitatively depend on the extra-dimensional size, 
but the ISCO and UPCO take various radii 
according to the sizes of mass and compactification.

It is inadvisable to apply this model to the Universe because the 
caged black hole has an electric charge and is justified only at $a/\sqrt{\mu}
\gg1$%
\footnote{
\begin{align}
\frac{a}{\sqrt{\mu}}
&\sim 10^{-4} \left(
\frac{M_{\odot}}{M_{\mathrm{grav}}}
\right)^{1/2}\left(
\frac{a}{0.1~\mathrm{mm}}
\right)^{1/2}
\sim 10^{23} \left(\frac{\mathrm{TeV}/c^2}{M_{\mathrm{grav}}
}
\right)^{1/2} \left(
\frac{a}{0.1~\mathrm{mm}}
\right)^{1/2}.
\end{align}}
(see, e.g., Ref.~\cite{Harmark:2002tr}).
Even if we applied it, we would find that the behavior at infinity is the same as in 4D, 
but for example, the ISCO radius takes a larger value $4.5 r_{\mathrm{g}}$ 
than the value $3r_{\mathrm{g}}$ we expect, 
where $r_{\mathrm{g}}$ is the Schwarzschild radius. 
Such behavior does not adequately represent the actual astrophysical situation.
If we consider the higher-dimensional Universe scenario in an astrophysical situation, 
then we may give a more realistic model by a squashed Kaluza-Klein black hole with a horizon expanding to the whole extra dimension, rather than a caged black hole. 
The interpretation of stable circular orbits proposed recently in the context of the AdS/CFT correspondence would also be interesting~\cite{Berenstein:2020vlp,Konoplya:2020ptx}.
These issues deserve further study.

\begin{acknowledgments}
This work was supported by the Grant-in-Aid 
for Early-Career Scientists~[JSPS KAKENHI Grant No.~JP19K14715 (T.I.)] and 
Grant-in-Aid for Scientific Research (C) [JSPS KAKENHI Grant No.~JP17K05452 (S.T.)]
from the Japan Society for the Promotion of Science. 
S.T. is also supported from Toyota Technological Institute Fund for
Research Promotion A.
\end{acknowledgments}
\appendix

\end{document}